# Gene-environment interplay and public policies

Dilnoza Muslimova
Erasmus School of Economics, Erasmus University Rotterdam, The Netherlands
E-mail: muslimova@ese.eur.nl

Cornelius A. Rietveld
Erasmus School of Economics, Erasmus University Rotterdam, The Netherlands
Tinbergen Institute, The Netherlands
E-mail: nrietveld@ese.eur.nl

**Abstract**

This *perspective* posits that gene-environment interplay (G×E) studies should be developed both theoretically and empirically to be of relevance to policy makers. On the theoretical front, this development is essential because the current literature lacks the integration of a clear framework capturing the various goals of public policies. Empirically, G×E models need to be further developed because the common way of modelling G×E effects fails to adequately capture the heterogeneous effects public policies may have along the distribution of genetic propensities (as captured by polygenic indices). We fill these gaps by proposing a policy classification for G×E research and by offering guidance on advancing the empirical modelling of policy-informative G×E interplay. While doing so, we provide a review of existing G×E studies on educational outcomes exploiting policy reforms or environments that could be targeted by public policy.

**Acknowledgments**

The authors gratefully acknowledge funding from the European Research Council (GEPSI 946647).

# Introduction

Genetic differences are one of the main contributors to inequalities in health and socio-economic status (Benjamin et al., 2012; Polderman et al., 2015; Bingley, Cappellari & Tatsiramos, 2023). Genes do not change over the life-course, but gene-environment (G×E) interplay research shows that environmental circumstances can either mitigate or exacerbate their impact on such inequalities (Domingue et al., 2020; Dias Pereira et al., 2022). With the reduction of these inequalities being of crucial importance for achieving a better and more sustainable future for all (United Nations, 2015), G×E research is often appraised in the social science genetics literature for its policy-relevance. The simultaneous modelling of both genetic and environmental factors in representative samples may inform policy makers which environmental circumstances to focus on to address inequalities stemming from genetic differences (Biroli et al., 2025; Herrera-Luis, Benke, Volk, Ladd-Acosta & Wojcik, 2024).[1]

There are several theoretical models on G×E interplay, which can be categorized into two groups (Liu and Guo, 2015; Mills, Barban & Tropf, 2020). Most prominent in the first group is the diathesis-stress model (also known as vulnerability or contextual triggering model), which hypothesizes that the genetic propensity for a trait remains unexpressed unless triggered by a specific environment. Mirroring this, the social control or social push model hypothesizes that specific environments dampen genetic effects. In the second set are the bioecological or social compensation model hypothesizing that genetic effects are maximized in certain environments and the differential susceptibility model allowing for both positive and negative responses of genetic effects to environments. Thus, while a discontinuity is central in the first set of models, the second set emphasizes proportionality in how genetic effects depend on environmental circumstances.

In line with this second set of models, the dominant hypothesis guiding G×E research in social science genetics is the so-called Scarr-Rowe hypothesis, predicting that genetic influences are maximized in favorable environments allowing a full expression of genetic propensities (Turkheimer, Harden, D'Onofrio & Gottesman, 2011; Ghirardi & Bernardi, 2023). For instance, the full expression of genetic propensities for educational attainment may be achieved in countries where generous student loan schemes allow highly talented students to pursue advanced degrees, irrespective of their financial situation. If this is particularly beneficial for those with a higher genetic propensity for education, which is not unlikely given the substantial returns on education (Carneiro, Heckman & Vytlacil, 2011), the introduction of such a policy may result in positive G×E complementarities and larger genetic inequalities.

However, policy makers can also purposely restrict the environments their citizens live in. Facilitated by the release of the UK Biobank data (Bycroft et al., 2018), an often-exploited environmental circumstance in G×E research is the 1972 "raising of schooling leave age" (RoSLA) reform in the United Kingdom (Barcellos et al., 2018; Davies et al., 2018), inhibiting individuals to drop out of school before the age of 16 (instead of 15 before this educational reform). Barcellos, Carvalho and Turley (2021) show that this reform particularly impacted those with a low genetic propensity for educational attainment by making them to stay in school longer. This result is in line with the implied inverse of the Scarr-Rowe hypothesis: a more restricted environment results in smaller genetic inequalities.

[1] As opposed to the modification of an individuals' genes. Recently, selection of embryos during IVF treatments has been advocated to bring genetic inheritances under some control (Turley et al., 2021). Still, measurement error in polygenic indices precludes precise individual-level predictions (Ding et al., 2022; Muslimova et al., 2023). Therefore, potential effects of such selection are, if any, tiny and come with potential unintended consequences such as selection for adverse traits and devaluation of certain traits. Apart from ethical considerations, statistical reasons thus already preclude the modification of genes as a target of public policies.

These two examples show that public policies aiming to improve educational outcomes may have differential effects on inequalities stemming from genes. The reason is that these policies differentially affect parts of the population, the first one mostly those with higher and the second one mostly those with lower genetic propensities for education. It is helpful here to draw a parallel with terminology used in the literature on human capital formation, by distinguishing reinforcement from compensation (e.g., Becker & Tomes, 1976; Behrman, Pollack & Taubman, 1982; Fan & Porter, 2018). There are public policies aiming to create a supportive environment that allows individuals and communities to thrive. The focus of such *reinforcing* policies is on removing barriers and providing opportunities to realize potential and thus to widen inequalities stemming from genes. However, there are also public policies aiming to address and rectify inequalities and disadvantages experienced by specific groups in society. The focus of such *compensatory* policies is on providing direct assistance to those who have been marginalized or disadvantaged.

The bifurcation also speaks to the discussion on the redistributive justice of inequalities stemming from genes. Focusing on educational attainment as a prime socio-economic characteristic, whereas most people would agree that high-ability children from poor backgrounds should be able to flourish in school (i.e., the principle of equality of opportunity; Roemer & Trannoy, 2015) there is vociferous disagreement about whether a tighter link between genotypes and educational attainment is considered positive or negative. Proponents consider a tight link between genotypes and outcomes as a signal of meritocracy (Rimfeld et al., 2018), whereas others point out that genotypes simply constitute yet another layer of inequality of opportunity since genes are inherited from the parents and do not represent individual effort (Harden, 2021; Kweon et al., 2020). Settling this debate is only possible when it is known where inequalities are concentrated in the distribution of genetic propensity.

In this *perspective*, we posit that G×E models should be both theoretically and empirically developed to be of relevance to policy makers. On the theoretical front, this is essential because the current literature does not clearly integrate policy frameworks to generate hypotheses on the impact of policy on genetic inequalities. Empirically, G×E model need to be further developed because the common way of modelling G×E effects fails to adequately capture the heterogeneous effects policies may have along the distribution of genetic propensity. We fill these gaps by proposing the policy classification developed by Lowi (1972) as a relevant lens for G×E research and by offering guidance on advancing the empirical modelling of policy-informative G×E interplay. While doing so, we provide a review of existing G×E studies on educational outcomes exploiting policy reforms or environmental interventions that could be targeted by policy. To ensure the actuality of the review, we focus on G×E studies using polygenic indices as a measure of genetic propensity for educational attainment. Polygenic indices summarize the genetic predisposition of an individual to a specific outcome based on the combined effect of common genetic variants in an independent training sample, and they are increasingly available in social science datasets (Alemu et al., 2025; Becker et al., 2021).

# A policy framework for G×E models

Understanding the types of policies and their goals is crucial when considering the implications of gene-environment interplay studies for inequalities in socio-economic outcomes stemming from genes. Policies can be classified in different ways. Some approaches, such as those proposed by Bemelmans-Videc, Rist, and Vedung (2017), focus on the nature of power. Other approaches, such as the framework offered by Lowi (1972), categorizes policies based on the intended impact on their beneficiaries, which makes it particularly informative for G×E analysis. Beyond its relevance, this classification has become a foundational framework in public policy analysis and is widely taught in public policy curricula (Sanders, 1990; Greitens & Joaquin, 2010). In this section, we discuss the classification of Lowi (1972) in more detail, map the gene-environment interplay literature on educational attainment to it, consider possible additional aspects of policy, and discuss implications for modelling gene-environment interplay.

In his 1972 article titled "Four Systems of Policy, Politics and Choice", Lowi identified four types of policies: distributive, redistributive, regulatory, and constituent (Lowi, 1972). Distributive policies involve the allocation of goods and services to individuals without significant costs to others, typically with the aim of promoting social or economic equity. Redistributive policies, on the other hand, reallocate resources from one group to another with the goal of reducing social and economic disparities. This difference can be illustrated by considering a study from the gene-environment interplay literature, which has investigated how teacher quality and quantity modify the effect of the genetic predisposition for education on educational attainment (Arold, Hufe & Stoeckli, 2025). A policy where investments are made in the quality of all teachers, would be considered distributive. However, if teacher quality investments are only realized in schools with a higher proportion of disadvantaged children to close achievement gaps, such a policy is clearly trying to shift resources and is hence, considered redistributive. Interestingly, in the first scenario, one would expect a positive interaction with genetic predisposition for education since the environment creates opportunities for everyone and, typically, those with a higher propensity for education retrieve higher returns from such environments (Muslimova et al., 2024). In the second scenario, however, the expected interaction would be negative if having a lower genetic propensity for education is correlated with lower socio-economic status (as for example shown by Barth et al. (2020)) since the policy is clearly trying to redistribute resources to less advantaged children.

Regulatory policies, in contrast, focus on controlling behaviours and practices through legal standards or procedures. For educational attainment, a relevant example is the Raising of the School Leaving Age (RoSLA), which requires students to remain in school until a certain age. RoSLA was implemented in the UK several times, most notably are the 1947 reform when the school leaving age was raised from 14 to 15 years and the 1972 reform when the school leaving age was raised once again by one year to 16 years. RoSLA was particularly successful in reducing educational inequalities, for example, years of schooling increased among all affected cohorts, but more so among those in the bottom tercile of the distribution of the polygenic index for education (Barcellos, Carvalho, and Turley, 2021).

Finally, constituent policies differ from distributive, redistributive, and regulatory policies in that they do not involve the allocation of resources or the regulation of behaviour of an individual citizen. Instead, constituent policies focus on the system, i.e., the organization of government structures and the values and beliefs that underpin a society. In the context of education, examples include the formation of educational agencies, commissions, and committees. These policies play a key role in shaping the governance and administration of educational systems and may reflect in the general quality of education.

Public policies tend to combine elements of distributive, redistributive, regulatory, and constituent nature. Ultimately, broadly defined, policy is a set of tools used by authorities to gain support of electorate and manage social change (Vedung, 1998), and in doing so, achieve a certain outcome. If one would simply view policy as an instrument that either encourages people to take action or prevents people from taking action, the degree of coercion becomes more relevant than the accurate classification of the policy (Pal, 2014). Hence, policies can be either coercive and sanction-based or rely on incentives and capacity building. Policies affecting inequalities in education can manifest different levels of coercion. For example, minimum leaving age policies such as RoSLA are coercive policies, forcing children to stay until school until a certain age. Subsidies for optional pre-school programs have a more incentivizing nature and are, hence, less coercive.

Coercive policies are designed to prevent certain or punish unlawful behavior, but in the context of inequalities policies with a high level of coercion appear to be mostly targeted to compensate less advantaged individuals. All the while, incentive-based policies or policies with low or no coercion, create opportunities for all, hence, they enable behaviors. In this case, the outcome depends on the take up of such opportunities. Theoretical G×E models have already alluded to the classification of environments into compensating and reinforcing independent of the public policy literature. However, bridging the two literatures is informative for building expectations on how policies would affect genetic inequalities. That is, in addition to the formal classifications and the level of coercion, policies can be classified as being compensating and/or reinforcing.

In Table 1, we summarize the above discussion by mapping the extant G×E literature on educational outcomes (see Appendix A for selection criteria) to reinforcing and compensatory policies and by assessing the degree of coercion. The assessment of policy type and the degree of coercion are based on the following two rules. Firstly, an environment in a study is categorized as reinforcing if it concerns a general improvement in the environment without a specific target group. An environment is categorized as compensatory if it is expected to target a disadvantaged subgroup of the population. Secondly, the degree of coercion (High/Low) reflects the extent to which the variation in the environment stems from the direct government intervention or regulation. Relatedly, although most of the presented environmental exposures are not policy interventions themselves directly, they are likely the result of policy interventions, which could not be observed due to data limitations. Thus, we present a summary of the gene-environment interplay literature, which either investigates policy interventions in education or environmental exposures relevant to policy makers.

Table 1 shows that reinforcing environments tend to interact positively with the genetic propensity for education—though not without exceptions. In addition, compensatory environments are generally marked by high coercion, whereas reinforcing environments appear less coercive. The review, however, indicates that these patterns are not systematic; the direction of G×E interactions remains context-dependent. Further, the majority of the studies presented in Table 1 investigate how the effect of the environments differs among individuals with a higher versus lower genetic propensity for education. Yet, it is also possible that some policies differentially affect those in the middle of this distribution. Although evidence regarding environments impacting particularly those in the middle of distribution of genetic propensities is hard to pinpoint from this review, this possibility cannot be ruled out without a model that explicitly allows for testing this. For instance, the impact of "one-size-fits-all" educational approaches may be largest for students with an average genetic propensity for education. Moreover, it is also important to acknowledge that studies of gene-environment interplay are susceptible to publication bias, as null or weak interactions stemming from conventional models often go unreported.

Given the diversity of policies and their varied effect on individuals along the distribution of genetic propensity for education, the current way of modelling G×E interactions – typically a simplified linear specification – appears increasingly inadequate. Historically, this linear framework has been adopted due to limited statistical power to detect subtle interactions. However, with the advent of larger datasets, there is now a clear imperative to model G×E interactions in a more informative way. In line with the discussion above, the next section outlines potential extensions to the traditional G×E framework.

***Table 1.*** *Gene-environment (G×E) interplay in educational outcomes in studies employing a polygenic index (PGI) for educational attainment (alphabetical order).*

| Study | Outcome | Environment | Sign G×E | Policy type | Coercion | Interpretation |
|---|---|---|---|---|---|---|
| Arold, Hufe & Stoeckli (2025) | Years of education | Teacher quality | Negative | Compensatory / Reinforcing | High | Higher quality schools mitigate educational inequalities stemming from genetic endowments and can be substitutes for genetic endowments. |
| Barban, Mills & Tropf (2017) | Years of education | Proportion (at census block level) of individuals with college degree, income below poverty level, unemployment rate, single mothers, median household income | Positive | Reinforcing | Low | Genetic effects are more pronounced in neighbourhoods with higher affluence as compared to deprived neighbourhoods. |
| Barcellos et al. (2021) | School leaving age, passing qualifications | Raising of Schooling Leaving Age (RoSLA) 1972 | Negative | Compensatory | High | The 1972 RoSLA reduces the gap in average school leaving age between individuals in the top and bottom tercile of the PGI distribution. |
| Biroli et al. (2025) | Test scores (Entry assessment, Key Stage 1-4) | School entry policy (being old-for-grade) | Positive / Negative | Compensatory | High | Entry assessment test results depict positive interaction between being old-for-grade, while Key Stage test results 1-4 negative, suggesting families increase inequalities stemming from genetic differences, while schools reduce them. |
| Cheesman et al. (2020) | Years of education | Home environment (non-adopted vs. adopted) | Negative | Reinforcing | Low | Educationally supportive home environments compensate inequalities in education driven by genetic differences. Specifically, adopted individuals in the lowest decile of the PGI distribution achieve significantly more education as compared to their non-adopted peers. |
| Cheesman et al. (2022) | School achievement (math, reading, English) | School | Negative | Compensatory / Reinforcing | High | School differences matter more for educational achievement of children with lower PGI scores as compared to their peers on the higher end of PGI distribution. |
| Conley et al. (2019) | Years of education | Birth weight | Mixed to null | Reinforcing | Low | No robust interactions identified. |
| Fletcher (2023) | Years of education | Birth-state-by-decade social mobility | Negative | Reinforcing | Low | The association of the PGI with educational attainment is weaker among children who grew up in states with high educational mobility. |
| Fraemke et al. (2024) | Years of education | German unification | Positive | Reinforcing | High | PGI associations with educational attainment are significantly stronger among East Germans after unification as compared to West Germans and to the period before unification. |
| Herd et al. (2019) | Years of education | Gender×cohort | Positive | Reinforcing | Low | The relationship between the PGI and education became stronger for women as they aged due to more opportunities from the 1970s onwards. Gender differences in the relationship between genetics and educational outcomes weakened with increased access to education for women. |
| Houmark (2025) | School achievement in grades 2-8 | Family income and education | Negative | Reinforcing | Low | PGI becomes more predictive of reading scores with time for children with lower socio-economic background, the effect does not change for children with higher socio-economic background; schooling has a compensating effect with more time passing by. |
| Isunget et al. (2021) | Math, reading tests | Parental education | Not different from zero | Reinforcing | Low | No evidence of Scarr-Rowe type of interactions (effect of genetic variation is supressed in lower socio-economic status environments) in countries with broader welfare policies. |
| Lin (2020) | Years of education | Parental education | Negative | Reinforcing | Low | Combined with the positive effect of parental education, higher educated parents are able to assure advantages in schooling for offspring in the lower end of the PGI distribution. |
| Liu & Clark (2022) | Years of education | Paternal incarceration (absence) | Positive | Reinforcing | High | The association between the PGI and educational attainment is lower among those who experienced paternal incarceration as compared to those who never did. |
| Muslimova et al. (2024) | Years of education | Birth order | Positive | Reinforcing | Low | Parental investments are more productive for children in the higher end of the PGI distribution. |
| Papageorge & Thom (2020) | Completed degree | Childhood socio-economic status | Positive | Reinforcing | Low | The association of the PGI with higher rates of college graduation is stronger for individuals who grew up in households of higher socio-economic status relative to those who grew up in poorer households. |
| Rimfeld et al. (2018) | Years of education | Soviet vs. post-Soviet era | Positive | Reinforcing | High | PGI is more predictive of educational attainment in the Post-Soviet period as compared to in the Soviet era. |
| Ronda et al. (2022) | Years of education, School achievement | Parental human capital, family resources, family stability, and parental mental health | Positive | Reinforcing | Low | Genetic influences on human capital formation are significantly smaller for children growing up in disadvantaged environments than for children growing up in normal circumstances. |
| Schmitz & Conley (2017) | Highest degree obtained, total years of college education, years of education | Vietnam lottery | Negative | Not applicable | High | Veterans in the bottom 50% of the PGI distribution completed fewer years of schooling. |
| Trejo et al. (2018) | Post-secondary education | School status as mean percentage of mothers with high school diploma, school stratification as Gini coefficient of parental education | Positive/Negative | Reinforcing | Low | The interaction is positive in the Wisconsin Longitudinal Study, but negative in AddHealth. |
| Uchikoshi & Conley (2021) | Mathematics and science tracking | Parental socio-economic status | Positive | Reinforcing | Low | Parental socio-economic status moderates the influence of the PGI on the tracking outcomes for math outcomes in $10^{th}$ grade. |
| Ujma et al. (2022) | Years of education, college degree | Cohort reflecting pre- and post-Communist regime | No robust interactions | Reinforcing | High | Insufficient statistical power to detect interactions. |
| Von Hinke & Sorensen (2023) | Years of education, fluid intelligence | London smog (absence) | Positive | Reinforcing | High | London smog has a negative effect on educational attainment for those in the top 50% of the PGI distribution. |

## An empirical framework for G×E models

The conventional empirical approach to modelling gene-environment interplay is (Biroli et al., 2025; Domingue et al., 2020):

$$Y_i = \alpha + \beta_G G_i + \beta_E E_i + \beta_{G\times E}(G_i \times E_i) + \epsilon_i, \quad (1)$$

where $Y_i$ is the outcome of interest, $G_i$ is a measure of genetic propensity typically measured by a polygenic index, and $E_i$ is an environmental shock or treatment. While we treat $G_i$ as a general measure of genetic propensity, we note that polygenic indices are typically constructed using the results of additive models from genome-wide association studies (Becker et al., 2021). Building on a less restrictive linear mixed model, Miao et al. (2025) show that such polygenic indices can still be used to empirically model gene-environment interplay as in Equation 1. Nevertheless, while this specification allows testing for the presence of linear interactions as illustrated in Figure 2 (panel **a**), it is not fully informative regarding which specific part of the polygenic index distribution is affected because the functional form only allows for an effect moderation along the entire distribution of *G*.

To allow for greater flexibility and to align better with a hypothesized conceptual G×E model, the sample is sometimes stratified by quantiles of the distribution of *G* to check for non-linearities. An important disadvantage of this approach is that the variance of *Y* may differ across the quantiles (Domingue et al., 2022; Purcell, 2002), making the effect sizes difficult to compare. Moreover, the number of quantiles in such stratifications is usually arbitrary. One way to remedy this would be to plot predictive margins of the outcome across different combinations of the polygenic index and the environment of choice as it is, for instance, done by Arold et al (2025, Figure 3). Still, this approach bypasses that classification into deciles of a polygenic index distribution comes with great imprecision due to measurement error in the polygenic indices (Ding et al., 2022; Muslimova et al., 2023).

With public policies potentially having differential effects along the PGI distribution, we thus need more flexibility than Equation 1 allows for. One straightforward way to extend the G×E estimation specification from Equation 1 that could capture the distributional effects of policies on genetic inequalities in a more flexible way is:

$$Y_i = \alpha + \gamma_G G_i + \gamma_E E_i + \gamma_{G\times E}(G_i \times E_i) + \gamma_{G^2} {G_i}^2 + \gamma_{G^2\times E}({G_i}^2 \times E_i) + \varepsilon_i, \quad (2)$$

where the addition of the squared term ${G_i}^2$ and its interaction with the treatment ${G_i}^2 \times E_i$ would allow testing for the distributional effects of the treatment $E_i$. Assuming $\gamma_G \neq 0, \gamma_E \neq 0, \gamma_{G\times E} \neq 0$, we can distinguish the following scenarios:

1. $\gamma_{G^2} = 0, \gamma_{G^2\times E} = 0$. Here, the effect of the polygenic index on the outcome differs for treatment and control groups, however, the difference is independent of the polygenic index (Figure 2, panel **a**). More precisely,
$$\frac{\partial Y}{\partial G} = \gamma_G + \gamma_{G\times E} E_i. \quad (3)$$
This scenario captures the typical G×E interplay specification (*cf.* Equation 1), and fits well with the bioecological, social compensation model, and the differential susceptibility model because of the proportionality in the response to the environment.

2. $\gamma_{G^2} \neq 0, \gamma_{G^2\times E} = 0$. This scenario captures the cases when the effect of the polygenic index on the outcome varies not only with the treatment but also with the polygenic index. Conceptually, this fits with the diathesis-stress model and the social control/push model because it allows for environment triggering a response in one part of the distribution of genetic propensity only. An attractive feature of this specification is that is allows identifying the point along the distribution of the polygenic index where the environment start playing a role. Instead of by ex-ante classification of individuals into quantiles, this point can be identified in a non-arbitrary manner by means of solving:

$$\frac{\partial Y}{\partial G} = \gamma_G + \gamma_{G\times E}E_i + 2\gamma_{G^2}G_i = 0, \tag{4}$$

$$\text{with optimum at } G_i = -\frac{\gamma_G + \gamma_{G\times E}E_i}{2\gamma_{G\times E}}. \tag{5}$$

The point where the environment starts playing a role may either lie inside or outside the relevant range of the polygenic index distribution. However, even if it lies outside the relevant range, it may meaningfully affect the shape of interaction effect within the relevant range, potentially contributing to a better fit between model and data than Equation 1 would allow for.

3. $\gamma_{G^2} \neq 0, \gamma_{G^2\times E} \neq 0.$ In this scenario, there is full flexibility in the interactive effect of the treatment. Over and above the second scenario, it can additionally capture cases when the tail ends of the PGI distribution are not affected, while the middle of the distribution is affected (Figure 2, panel **d**). If the relationship between the polygenic index and the outcome appears to be quadratic at the baseline ($\gamma_{G^2} \neq 0$), it is relatively straightforward to hypothesize such an interaction. The change in the outcome due to a unit change in the polygenic index is then a function of the polygenic index itself:

$$\frac{\partial Y}{\partial G} = \gamma_G + \gamma_{G\times E}E_i + 2\gamma_{G^2}G_i + 2\gamma_{G^2\times E}(G_i \times E_i). \tag{6}$$

Although there is no specific theoretical G×E model this scenario can be associated with, the flexibility of this functional form can prove to be of added value in empirical analyses as it also allows for greater precision to capture the models pertaining to the second scenario (Figure 2, panels **b**,**c**).

Extending G×E models to include non-linear interactions as in Equation 2 increases the model complexity. To assess the feasibility of this approach, we conducted simulation-based power analyses. We find that a sample of at least 2,000 individuals is needed to detect $\gamma_{G^2}$ with 80% power, and a sample of at least 10,000 individuals to detect $\gamma_{G^2\times E}$. Still, these results show that the empirical framework proposed here is within reach of most existing datasets. A detailed description of the power calculations is provided in Appendix B.

***Figure 2.*** *Different forms of G×E interactions. Panel **a** captures the common approach to model G×E interplay, panels **b** and **c** capture situations in which the environment moderates the genetic effect in the left or right tail of the distribution of genetic propensity, respectively: Panel **b** illustrates an interaction with a compensatory policy environment, while panel **c** illustrates an interaction with a reinforcing environment. Panel **d** illustrates a situation in which the environment affects the genetic effect in the middle of the distribution of genetic propensity for an outcome of interest.*

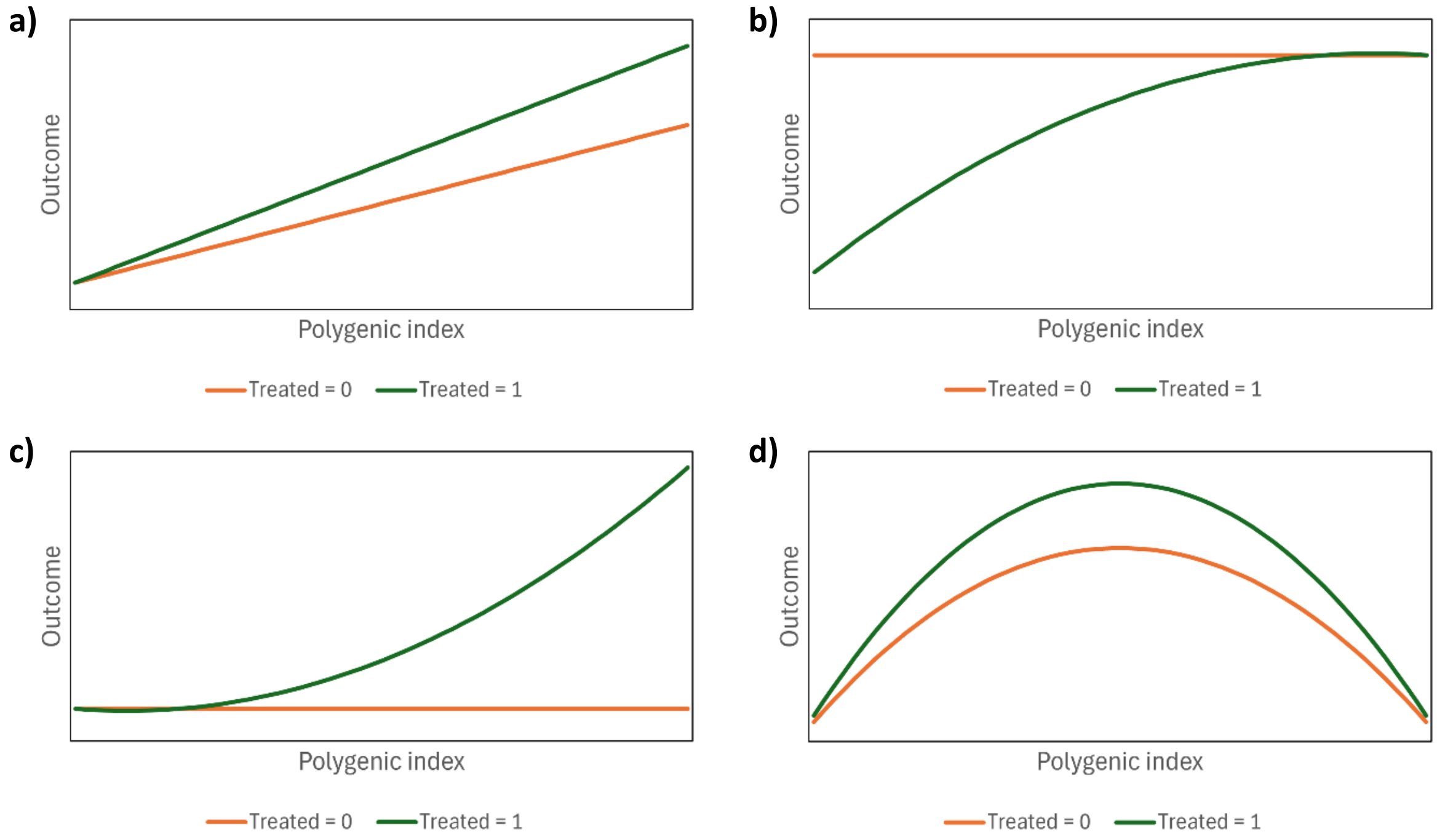

# Conclusion

Understanding the policy implications of gene-environment (G×E) interactions—and, consequently, the role of policy in shaping inequalities stemming from genetic differences—requires addressing two key gaps. First, existing research often lacks engagement with public policy frameworks that distinguish between policy types and objectives. Second, common empirical approaches to G×E interactions remain constrained by assumptions about their directional effects, e.g. strictly positive or negative interplay, limiting the ability to capture more complex patterns.

In this *perspective*, we bridge foundational public policy classifications with theoretical models of gene-environment interplay and map the existing empirical evidence on gene-environment interactions in educational outcomes onto the policy literature. Specifically, we build on Lowi's (1972) seminal taxonomy—which categorizes policies as distributive, redistributive, regulatory, or constituent—combined with the dimension of coercion to assess whether policies primarily reinforce or compensate for individual differences. This framework can complement the existing theoretical modelling of G×E interactions by drawing researchers' attention to the ex-ante intentions of a policy or an environmental change when hypothesizing about the interactions and empirically estimating them. Building on this conceptual foundation, a natural expectation is that reinforcing policies amplify genetic predispositions for education, whereas compensatory policies dampen them. Additionally, more coercive policies would have more widespread effects on the target population than less coercive policies due to enforcement. However, our review suggests that such patterns are not systematic. Instead, the sign of G×E interactions appears to be context-dependent. At the same time, we underscore that while the gene-environment literature proliferated with the availability of the polygenic scores, evidence on robust causal interactions is still very limited (Biroli et al., 2025).

We further highlight the constraints of current empirical models in testing G×E interactions, which often impose rigid assumptions about the direction and form of interactions. Ghirardi and Bernardi (2024) recently demonstrated that the sign of interactions between polygenic indices and socio-economic status can shift when moving from less selective to more selective educational outcomes (e.g., years of schooling vs. PhD attainment). Similarly, Domingue et al. (2020) show that transitioning from continuous to binary outcome measures can yield inconsistent G×E findings. In addition, these authors underscore that the lack of consideration of how environmental factors alter outcome distributions may lead to spurious interactions. We propose a flexible framework that allows for interactions beyond strictly positive or negative effects. By allowing the interaction term to depend on the polygenic index in a more nuanced way, it allows for testing hypotheses derived from Lowi's (1972) four policy types with varying degrees of coercion.

Building on these insights, we argue for incorporating public policy classifications to qualify environmental exposures and propose methodological refinements to enhance empirical flexibility to study G×E interplay. Future research should not only advance causal identification in G×E interactions (Biroli et al., 2025) but also deepen our understanding of how public policy shapes inequalities in education, health, and other important life outcomes, stemming from genetic differences. To this end, we join earlier calls for the collection of genotyped samples that are fully representative of the relevant populations (Mills & Rahal, 2020) to understand why certain policies work or not work and for whom.

# Appendix A

Appendix A summarizes the details of the literature selection for Table 1. The following selection criteria were applied using a snowballing approach:

1. Search terms: educational attainment, years of education, educational achievement, exam scores, test scores, gene-environment interplay/interaction, nature-nurture interplay/interaction, polygenic index for educational attainment
2. Inclusion criteria:
    - Studies are published or have an online pre-print;
    - Educational outcomes are measured as years of education or as exam/test scores;
    - Genetic measure employed is a polygenic index for educational attainment;
    - Environmental exposure is shown to affect educational attainment and performance in non-G×E studies.
3. Exclusion criteria:
    - No candidate gene-studies;
    - Only G×E in educational outcomes;
    - Only socio-economically relevant environments.

# Appendix B

To assess the feasibility of estimating the extended model from Equation 2, we conducted simulation-based power calculations. Our simulations assume a standardized polygenic index, a binary environmental exposure (50% prevalence), and effect sizes for nonlinear terms expressed as fractions of the corresponding linear effects. Following the assumptions made by Biroli et al. (2025) regarding their G×E model on the effect of a school entry age policy on educational outcomes, we assume $\gamma_G$ = 0.259 and $\gamma_E$ = 0.90 (Entry Assessment test score), and take $\gamma_{G\times E}$ to equal one-quarter of $\beta_E$. Additionally, we assume that $\gamma_{G^2}$ is one-quarter of $\gamma_G$, and $\gamma_{G^2\times E}$ to be the same proportion of $\gamma_{G\times E}$. The simulations are based on 1,000 runs, and we evaluate power at a significance level of 0.05.

Figure B1 shows that the results differs for $\gamma_{G^2}$ and $\gamma_{G^2\times E}$, with a required sample size of approximately 2,000 and 10,000, respectively, to be detected with 80% power. These findings suggest that the empirical framework proposed here is within reach of most existing datasets and scalable to more complex environmental exposures.

***Figure B1.*** *Statistical power to detect $\gamma_{G^2}$ and $\gamma_{G^2\times E}$ as function of sample size.*

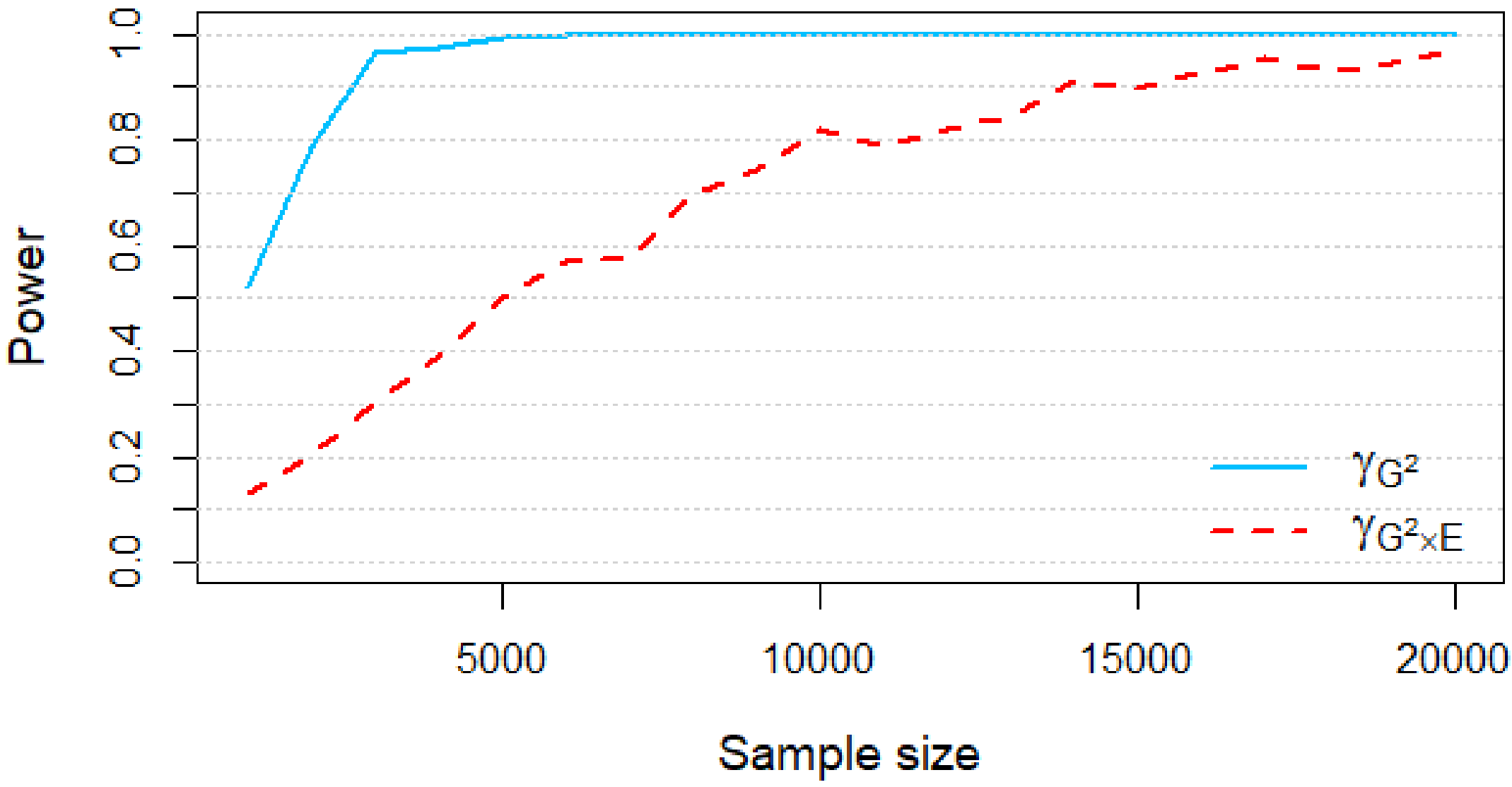